\documentclass[preprint,12pt]{elsarticle}

\usepackage{subcaption,graphicx}
\usepackage{dcolumn}
\usepackage{bm}
\usepackage{amsfonts}
\usepackage{gensymb}
\usepackage{amsmath}
\usepackage{amssymb}
\usepackage{latexsym}
\usepackage{textcomp}
\usepackage{color}
\usepackage{tikz}
\biboptions{numbers,sort&compress}

\PassOptionsToPackage{normalem}{ulem}
\usepackage{ulem}
\providecolor{added}{rgb}{0,0,1}
\providecolor{deleted}{rgb}{1,0,0}

\journal{Applied Surface Science}

\begin{document}
\def\myfrac#1#2{\frac{\displaystyle #1}{\displaystyle #2}}

\begin{frontmatter}

\title{Simultaneous nanopatterning and reduction of graphene oxide by femtosecond laser pulses} 

\author{Maren Kasischke}
\ead{kasischke@lat.rub.de}
\address{Applied Laser Technologies, Ruhr-Universit\"at Bochum,
Universit\"atsstra\ss e~150, 44801 Bochum, Germany}

\author{Stella Maragkaki}
\address{Applied Laser Technologies, Ruhr-Universit\"at Bochum,
Universit\"atsstra\ss e~150, 44801 Bochum, Germany}

\author{Sergej Volz}
\address{Applied Laser Technologies, Ruhr-Universit\"at Bochum,
Universit\"atsstra\ss e~150, 44801 Bochum, Germany}

\author{Andreas Ostendorf}
\address{Applied Laser Technologies, Ruhr-Universit\"at Bochum,
Universit\"atsstra\ss e~150, 44801 Bochum, Germany}

\author{Evgeny L. Gurevich}
\ead{gurevich@lat.rub.de}
\address{Applied Laser Technologies, Ruhr-Universit\"at Bochum,
Universit\"atsstra\ss e~150, 44801 Bochum, Germany}

\date{\today}

\begin{abstract}

This paper presents a novel one-step method for the periodical nanopatterning and reduction of graphene oxide (GO).  Self-organized periodic structures of reduced graphene oxide (rGO) appear on GO surfaces upon processing with a femtosecond laser at fluences slightly higher than the fluence needed for reduction of the GO. This indicates that the periodic pattern is formed either \textit{simultaneously with} or \textit{due to} the reduction of the GO. The laser-induced reduction of GO was identified by sheet resistance measurements, Raman and X-ray photoelectron spectroscopy. This fast and simple method to both reduce and periodically structure GO offers a variety of possible applications in printed and flexible electronics. 

\end{abstract}


\end{frontmatter}


\section{Introduction}
Graphene has received much attention in recent years due to its extraordinary physical properties \cite{Novoselov.2005}, which offer a wide range of possible applications \cite{Ferrari.2015}. Graphene oxide (GO) is a graphene-based material, which can be homogeneously dispersed in solvents and therefore provides an attractive option for cost-effective, large-scale production \cite{Zhou2010}. Reduction of GO has been proven to be an effective method toward the development of graphene-modified materials which have many possible uses, such as transparent electrodes, sensors, energy devices and biological chips \cite{Zhou2010,Hong2012,Trusovas.2016b}. Usually, the processing is done in two separate steps, which are reduction of the GO and patterning of the layer. However, a low-cost and fast processed production and patterning of these graphene-like materials in sub-micrometer scale is very challenging.

So far, patterned structures on graphene and GO have been achieved through several physical and chemical techniques \cite{Hong2012}, which can be classified into lithography, such as mask lithography \cite{Pang2009}, soft-lithography including transfer printing \cite{Allen2009,Jiang2014}, and direct laser writing \cite{FattTeoh.2012,Sokolov.2013,Yung.2013,Strong.2012b,Zhang.2010} including periodic structuring of multi-layer graphene \cite{beltaos2014femtosecond} and two-beam laser interference reduction and patterning of GO \cite{liu2016surface,guo2012two}.  These methods, however, are either based on patterning graphene after or before its production. 

One way to both pattern and reduce GO with a single technique is to apply laser processing. In particular, periodical patterning of graphene and rGO surfaces have a great potential in functionalizing the surface, e.g. for energy storage applications \cite{lin2014laser,pfleging2014new}, and in creating the periodicity required for plasmonic applications, like tunable terahertz metamaterials \cite{ju2011graphene} or tunable optical filters or broad-band modulators \cite{gao2012excitation}. Periodical structures in rGO have been achieved by two-beam laser interference for its application as sensing device \cite{guo2012two}, where however only the areas exposed with constructive interference were fully reduced. This paper presents a novel approach to simultaneously reduce and periodically pattern GO in one single step without the use of additional optical configurations or any extra chemical etching. In order to achieve a pattern on the surface, laser-induced periodic surface structures (LIPSS) \cite{BonseRev}, are employed. Generation of LIPSS is a single-step process, in which a periodic nanostructure appears upon the surface illumination with intense single or multiple laser pulses. LIPSS usually have a periodicity close to the laser wavelength and an orientation perpendicular to the incident light polarization \cite{BonseRev}. The physical background of the pattern formation is not clear and the main approaches are based either on the optical interference effect between the incident laser light and the surface scattered waves \cite{Sipe}, or on self-organisation processes on the surface \cite{Costache,varlamova2014genesis}. Usually these theories assume no changes in the chemical composition of the sample during the LIPSS formation. In this paper we report on simultaneous reduction and periodic patterning of GO, which can accelerate and simplify production of graphene nanostructures as well as provide new information for the LIPSS formation mechanisms.

\section{Experimental Setup}
In this study, commercially acquired graphene oxide (GO) in an aqueous solution with a concentration of 4~mg$/$mL (Graphenea S.A., Spain) is used and spin coated onto SiO$\textsubscript{2}$ substrates (Alineason Materials Technology GmbH, Germany). The spin coating process is repeated four times with a spin coating velocity of 5000 RPM and a subsequent heating at 95~$\degree$C for 10~min. The SiO$_{2}$ substrates had a polished surface with R$_a$$<$1\,nm. The thickness of the GO layer is $23.5$\,$\pm\,4.3\,$nm, measured with a laser scanning microscope VK-X200 (Keyence Corp., Japan).
For laser irradiation of GO a fiber-rod amplified femtosecond laser (Tangerine, Amplitude Systemes, France, average power $P=20$\,W, wavelength $\lambda=1030$\,nm, repetition rate $f=2$\,MHz, pulse duration $\tau_p\approx 280$\,fs) is employed. The laser beam is directed and focused onto the sample  by a galvanometer scanner and a 63\,mm f-theta objective. The calculated beam radius at beam waist is $\omega_0 = 10\,\mu$m taking M\textsuperscript{2}$= 1.05\,$ and a near Gaussian beam profile into account. Energy density is adjusted by a combination of a $\lambda \backslash 2$ plate and a polarizing beam splitter. During laser processing the sample was placed into a flow chamber with an argon flow of 10~L$/$min for inert atmosphere while processing in order to exclude any chemical reaction with the environment, which can also influence the LIPSS formation \cite{Zimmer2008}.
In order to analyze the influence of the peak fluence $F$ on GO modification it was varied in a range of $5.6$\,mJ/cm$^2$ to $56$\,mJ/cm$^2$ by varying the average output power of the laser. Also, the number of pulses per spot $N$ was varied from $10^2$\,pulses/$\mu$m to $2\times10^4$\,pulses/$\mu$m by controlling the scanning speed of the laser beam on the sample surface.

\section{Characterization Methods}
After laser processing the samples are analyzed by means of optical and scanning electron microscopy (SEM). In order to ascertain GO reduction Raman and X-ray photoelecton spectroscopy (XPS) are employed. Raman spectroscopy (inVia, renishaw GmbH, Germany) is performed with $\lambda$=532~nm and focused onto the sample with a 50$\times$objective. For avoiding sample damage a maximum power of 3.6~mW is employed while measuring. Data analysis is performed after baseline subtraction and smoothing of spectra. XPS is carried out with a Versaprobe spectrometer (Physical Electronics, USA) with monocromatized AlK$_\alpha$ radiation of 1486.6~eV. The UNIFIT2013 software is used for analyzing the survey and the high resolution measurements of the C1s and O1s peaks. 
Atomic force microscopy (AFM, Nanoscope 5, Bruker Corp. USA) is employed at ScanAsyst mode, which is based on PeakForce Tapping technology, in order to analyze the topography of the sample.

Sheet resistance is measured following the method of van der Pauw \cite{vanderPauwLeoJ.1958,Schroder.2006}, which uses a four point probe set up. For this purpose bridge structures are reduced by laser irradiation into the GO coating using different fluences and subsequently four 1$\times$1~mm silver electrodes are sputtered onto the patterns in order to avoid damaging the rGO layer with the measuring tips (see SI, Figure S1, for details). 

\section{Graphene Oxide Reduction}
\subsection{X-ray Photoelectron Spectroscopy}
\begin{figure}[h!]
\centering
\includegraphics[height=8 cm]{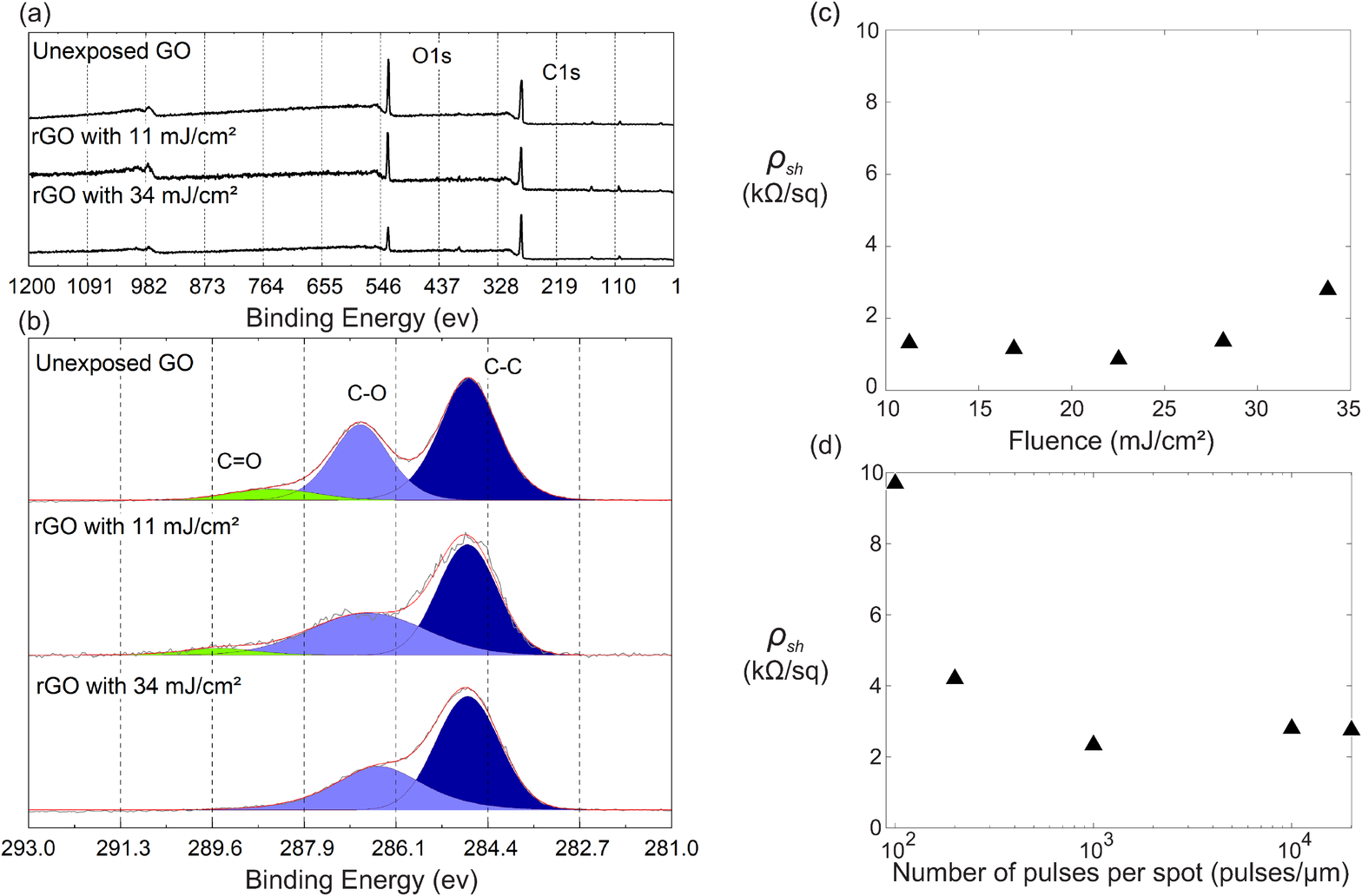}
\caption{XPS of rGO show a decrease in oxygen bonded carbon content compared to unexposed GO in both (a) survey where data is normalized to C1s peak (285.5 eV) and (b) high resolution spectra of C1s. Sheet resistance of rGO as a function of (c) incident laser fluence (at $10^4$ number of pulses per spot) and (d) number of pulses per spot (at 34\,mJ/cm$^2$) in a semilog plot.}
\label{XPSandRsh}
\end{figure}

 The analysis of the surface chemistry of GO after laser exposure by means of XPS, shown in Figure~\ref{XPSandRsh}~(a)~and~(b), indicate that the oxygen bonded carbon is significantly reduced, especially when applying higher laser fluence. The C1s high resolution spectra of the unexposed GO and the rGO areas are fitted using three components corresponding to C-C bonds (284 eV), epoxy -C-O-C (286 eV) and carbonyl -C=O (289 eV) groups. In Figure~\ref{XPSandRsh}~(b) a decrease of the oxygen bonded carbon of the laser exposed GO area compared to the pristine GO is detected. This can be attributed to the removal of epoxy- and carbonyl groups of the GO basal plane. Comparing the XPS spectra of the GO area exposed with low laser fluence ($11\,$mJ/cm$^2$) with the GO area exposed at higher fluence ($34\,$mJ/cm$^2$) the reduction of the oxygen content is more pronounced for the latter. 
 
\subsection{Sheet Resistance}
A substantial reduction of sheet resistance $\rho_{sh}$ was measured of the rGO layers compared to the unexposed GO. In literature a variety of values for the sheet resistance of unexposed GO has been documented ranging from  $10^{6}$~to~$10^{10}$~$\Omega/$sq \cite{Kymakis.2013,Sokolov.2013,Strong.2012b}, which may be due to the material  and the measurement method used. The results are plotted in Figure~\ref{XPSandRsh}~(c)~and~(d). The sheet resistance of areas irradiated with $10^4$ pulses per spot and fluences ranging from 11 to 34\,mJ/cm$^2$ ranks from 0.86$\pm0.01$~k$\Omega$/sq to 2.80$\pm0.01$~k$\Omega$/sq, see Figure~\ref{XPSandRsh}~(c). The increase of number of pulses at a set laser fluence has a more notable influence on the sheet resistance, which at $10^2\,$pulses per spot is as high as 9.7$\pm0.1$~k$\Omega/$sq while at $2\times10^4$ number of pulses per spot $\rho_{sh}=2.8\pm0.01$~k$\Omega$/sq, see Figure~\ref{XPSandRsh}~(d).

\subsection{Raman Spectroscopy}
Raman spectroscopy allows the analysis of the structure of the carbon network of the graphene basal plane present in the GO flake \cite{Trusovas.2016b}. The three most prominent peaks (D band at 1350~cm$^{-1}$, G band at 1600~cm$^{-1}$ and 2D band at 2698~cm$^{-1}$) of the Raman spectra are characteristic for graphene related materials \cite{Ferrari.2013}. The D band appears due to the breathing mode of the carbon atom rings and is activated by disorder of the ideal graphite network, where D stands for 'disorder' \cite{Ferrari.2013}. In particular, GO based samples exhibit a high D band intensity due to the sp$^3$ hybridized graphite oxide carbon bonds \cite{Sokolov.2013}. The G band appears due to in-plane vibration of the high-frequency $E_{2g}$ mode and features sp$^2$ hybridized carbon atoms in the network \cite{Ferrari.2013}. The 2D peak is the second order frequency of the D band and can indicate the presence of turbostatic graphite comprised of multiple graphene sheets which have an arbitrary orientation in the layer \cite{Sokolov.2013}.  
\begin{figure}[h]
\centering
\includegraphics[height=5 cm]{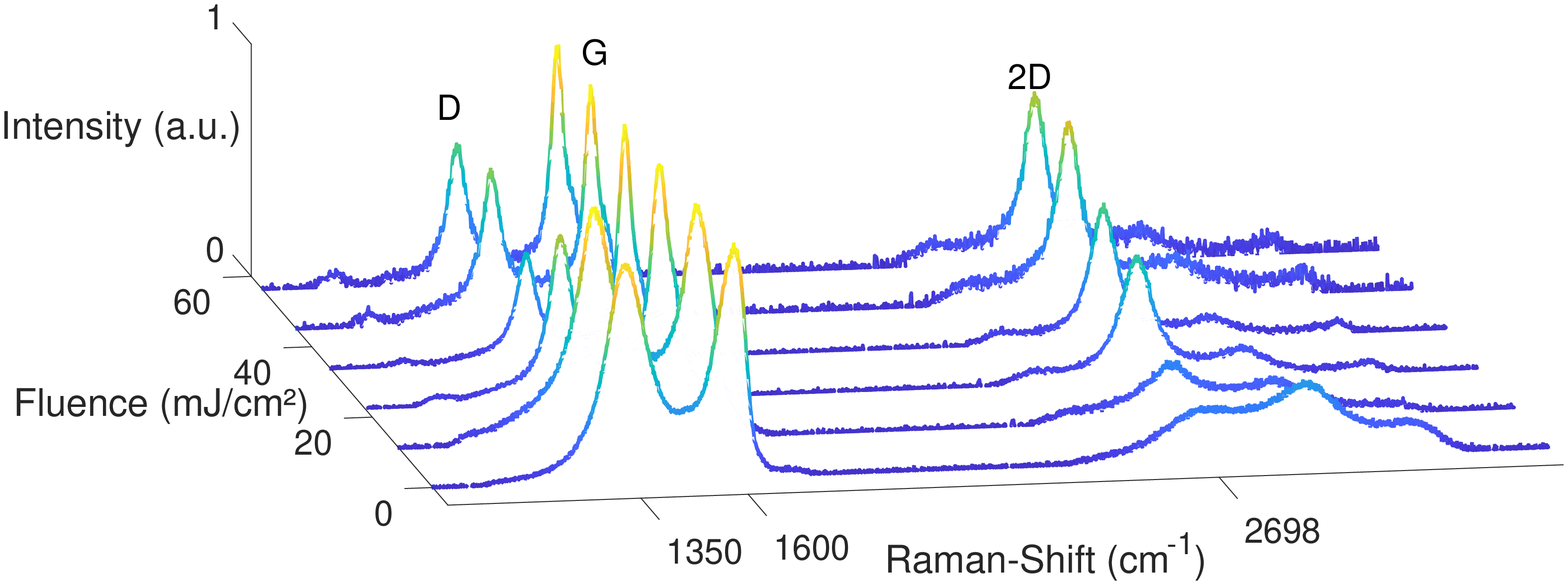}
\caption{Raman spectra of unexposed GO and reduced GO with increasing femtosecond laser fluence (at $10^4$ number of pulses per spot) show a significant increase of 2D peak and D peak decrease, indicating reduction of GO for $F\gtrsim11$\,mJ/cm$^2$}.
\label{Raman-GO}
\end{figure}

The Raman spectra of GO and rGO normalized on the amplitude of the G peak in dependency of the laser fluence applied are illustrated in Figure~\ref{Raman-GO}. Compared to the Raman spectrum of the unexposed GO the spectra of the rGO have a more prominent 2D peak and a lower D band. Whereby these features clearly intensify when increasing fluence. At fluences $F\gtrsim 35$\,mJ/cm$^2$ the D band intensity increases again. Similar results were found when increasing the overlapping rate of pulses (data shown in SI, Figure S2).

These changes in the Raman spectra indicate a fs-laser induced sp$^3$ to sp$^2$ structural reorganization in the carbon network. Previous studies \cite{Kymakis.2013,Arul.2016b}, where results of GO reduction with pulse durations within the fs and ns range were compared, indicate that for such structural transition of sp$^3$ to sp$^2$ graphene-like structures heat deposition into the material is necessary. In this study, such transformation was achieved with sufficient fluence ($F>11\,$mJ/cm$^2$) and number of pulses per spot ($N>10^3$\,pulses/$\mu$m) in inert atmosphere, whereby the mechanism for structural reorganization of GO may be most probably attributed to thermal energy deposition into the material. Indeed, for high repetition rate ($f=2$\,MHz) and low heat diffusion in the substrate, the layer temperature does not cool down between the laser pulses and heat accumulation happens \cite{Weber2014}. However, multiphoton mechanism of  GO reduction \cite{liu2016surface} and monolayer graphene photo-oxidation \cite{Aumanen.2015,bobrinetskiy2015patterned} is also discussed in literature.

Comparing the Raman spectrum of the area irradiated at a low fluence ($11\,$mJ/cm$^2$) with its corresponding XPS spectrum in Figure~\ref{XPSandRsh}~(b) shows that, even though barely no increase of the 2D band was detected, a reduction of the oxygen bonded carbon content was indeed measured with XPS. These results suggest that selective removal of oxygen groups occurs while the carbon network order is essentially maintained. The ultrafast laser pulse duration is shorter than the process of energy transfer to the lattice itself, which enables this selective oxygen removal, comparable to the results of \cite{Kymakis.2014}. In contrast, at higher fluence ($34\,$mJ/cm$^2$) both oxygen content reduction (see Figure~\ref{XPSandRsh}~(b)) and structural reorganization of the carbon lattice (indicated by the emerging 2D peak in the Raman spectrum) were measured. Hereby the high number of pulses per spot and the required fluence may cause a photothermal effect, resulting in structural transformation of sp$^2$ to sp$^3$ hybridized carbon, as discussed above.

\begin{figure}[h]
\centering
\includegraphics[width=0.8\linewidth]{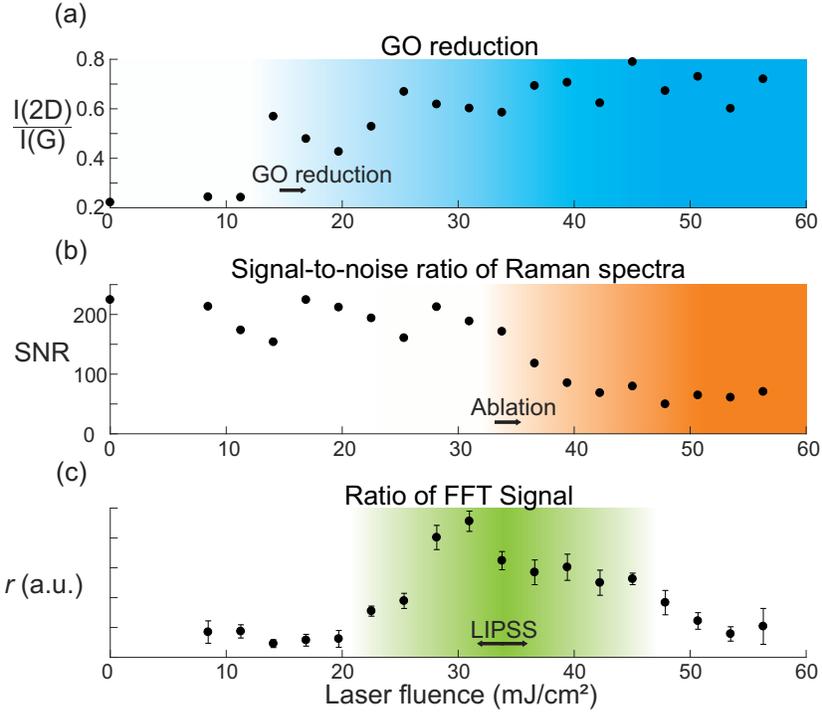}
\caption{Qualitative analysis of Raman spectra in (a) illustrates that GO reduction occurs at fluences above $F_r\approx11\,$mJ/cm$^2$, while (b) ablation of the rGO is increased at fluences above $F_a\approx35\,$mJ/cm$^2$. (c) The ratio $r$ calculated with 2D-FFT analysis of rGO images show LIPSS formation within a laser fluence range of $F_{l}\approx22.5 - 48\,$mJ/cm$^2$.}
\label{RamanLIPSS}
\end{figure}

The ratio between the intensity of 2D and G peaks in the Raman spectrum can be taken for qualitative description of the GO reduction \cite{Trusovas.2016b}. This ratio is plotted in Figure~\ref{RamanLIPSS}~(a) as a function of incident laser fluence. For $F<11\,$mJ/cm$^2$ the 2D peak is negligible, but above this value it starts growing and saturates, which means that for the laser-induced graphene reduction, the incident fluence should overcome a certain threshold value $F_r\approx11\,$mJ/cm$^2$. Simultaneously one can monitor the rGO ablation process by the ratio between the amplitude of the G peak and the standard deviation of the noise in the Raman spectrum measured between 1870 and 2300\,cm$^{-1}$, see Figure~\ref{RamanLIPSS}~(b). This ratio is constant for $F\lesssim 35\,$mJ/cm$^2$ and starts falling after this second threshold, namely $F_a\approx35\,$mJ/cm$^2$, indicating ablation of the rGO.

\subsection{Laser Induced Periodic Surface Structures}

\begin{figure}[h]
\centering
\includegraphics[height=9 cm]{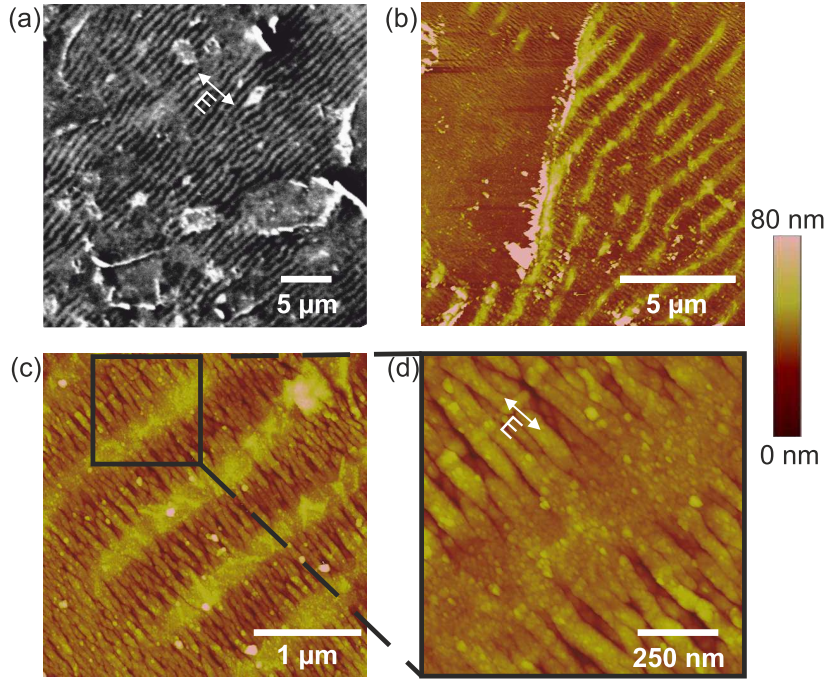}
\caption{(a) SEM and (b) AFM images of periodic surface structures, or LSFL,
with an average period of $800\,$nm, a height of $28\,$nm and an orientation perpendicular to the laser polarization as indicated by the white arrow on top in (a). (c) and (d) AFM images with higher magnification show both LSFL and HSFL. Latter have a periodicity of $110\,$nm and an orientation parallel to the laser light polarization, pointed out by white arrow in (d). Sample exposed with a laser fluence of $34\,$mJ/cm$^2$ and 10$^4$ number of pulses per spot.}
\label{REM-GO}
\end{figure}

For fluences slightly above $F_r$ an unexpected phenomenon appears: As can be seen in Figure~\ref{REM-GO}, laser-induced periodic surface structures arise on the laser-treated GO surface. These structures feature the same peculiarities as in other materials \cite{BonseRev}: the LIPSS consist of two periodic substructures with perpendicular orientation and different periods. The long-wavelength ripples, see Figures~\ref{REM-GO}~(a)~and~(b), are perpendicular to the polarization of the laser light and usually referred to as LSFL (low spatial frequency LIPSS).
The periodicity of these structures is approximately $\Lambda\approx0.80\,\mu$m, i.e., comparable to the incident light wavelength $(\Lambda\approx\lambda/1.3)$, as it is common for multi-pulse LIPSS \cite{BonseRev}.
The second substructure, see Figures~\ref{REM-GO}~(c) and (d), is parallel to the polarization and is characterized by a much smaller periodicity $\Lambda\approx0.11\,\mu$m which is about nine times smaller than the laser wavelength $(\Lambda\approx\lambda/9)$.
These structures with lower periodicity are known as HSFL (high spatial frequency LIPSS). HSFL usually appear at the contour of the laser exposed area due to the low laser intensity in this area when the beam profile intensity distribution is of Gaussian nature \cite{BonseGaussLIPSS} or due to presence of initial tranches on the surface \cite{Nathala2015}, whereas in this study these HSFL are nested in between the LSFL (see Figure~\ref{REM-GO} (c)). The period of the HSFL observed in this study is in accordance with the observation of LIPSS in multi-layer graphene produced by micromechanical exfoliation technique reported by Beltaos et al. \cite{beltaos2014femtosecond}.

The amplitude of the LSFL was measured to be $d=28\,\pm8.4\,$nm indicating, that a redistribution of the rGO layer takes place since it is in the range of the measured film thickness. In addition, Figures~\ref{REM-GO}~(a)~and~(b) show that at some parts the GO coating has partially peeled off and no periodic structures are present. These may have been areas with higher GO flake concentration, which have different thermal conductivity and optical properties than the rest of the coating and therefore detach from the layer upon laser irradiation. The distribution of these GO flakes in the layer is random and cannot be controlled when spin coating the GO dispersion.

\begin{figure}[h]
\centering
\includegraphics[width=\linewidth]{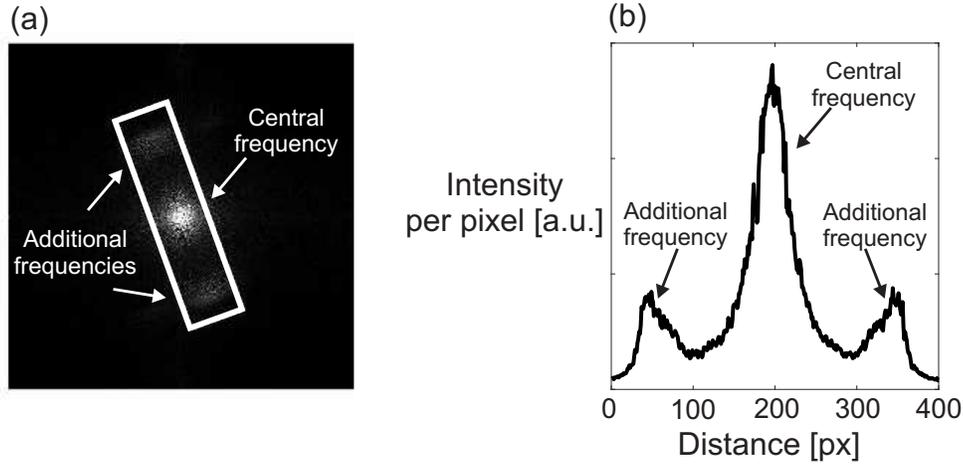}
\caption{Example of (a) 2D-FFT spectrum and (b) mean of 100 profiles taken at the white rectangle marked in (a) of rGO sample exposed at $31\,$mJ/cm$^2$, where the highest amount of periodic surface structures was observed.}
\label{FFT-example}
\end{figure}

The area where the LIPSS can be clearly detected depends on the laser fluence applied. In order to quantify this dependency an image analysis based on 2D fast fourier transformation (2D-FFT) was carried out. The presence of periodic structures is reflected in the  spectrum as additional peaks beside the central frequency (corresponding to the constant surface profile) \cite{gnilitskyi2017high}. The amplitudes of these additional peaks reflect the amount of the periodic structures present in the analyzed image. A profile of the intensity of the region of interest was calculated as the mean of 100 profiles of the marked region (see white rectangle in Figure~\ref{FFT-example}~(a) and mean of profiles in \ref{FFT-example}~(b)). The position and number of the taken profiles was the same for all analyzed images. Subsequently the ratio $r$ between the amplitudes of these peaks was plotted in Figure~\ref{RamanLIPSS}~(c) to quantify the efficiency of LIPSS formation at different fluences.

In the results of the image analysis (see Figure~\ref{RamanLIPSS}~(c)) three main domains can be identified. At lower fluences of the incident laser light below approximately 20\,mJ/cm$^2$ no, or barely any, periodic structures are observed, while at higher fluences $F_l\approx22.5 - 48\,$mJ/cm$^2$ these structures are present. At even higher fluences ($F\gtrsim50.5\,$mJ/cm$^2$) the periodic structures become scarcer until they disappear completely. Comparing the dependencies in Figure~\ref{RamanLIPSS}~(b) and Figure~\ref{RamanLIPSS}~(c), it can be supposed that the LIPSS disappear at higher fluences because the rGO layer is ablated.

\section{Discussion}

Three possible scenarios can be suggested to describe the appearance of the periodic surface structures on the surface upon femtosecond laser irradiation in our experiments:
\begin{enumerate}
\item The periodic structures are formed on the GO surface by the first laser pulses before the reduction, and then the GO layer with periodically modulated thickness is reduced by the following laser pulses.
\item First the GO is reduced and the LIPSS are formed in the conductive rGO.
\item Both reduction and LIPSS formation happen simultaneously.
\end{enumerate}
The first scenario (LIPSS formation before reduction) seems to be less probable, because the threshold for the reduction is below the threshold of the LIPSS formation, compare Figures~\ref{RamanLIPSS} (a) and (c). Indeed, GO reduction is observed for laser fluences above $F_r\approx11\,$mJ/cm$^2$, while the range of laser fluences needed for LIPSS formation is $F_l\approx22.5 - 48\,$mJ/cm$^2$.

The second scenario (first reduction, then LIPSS formation) would agree with the theory of the LIPSS formation, which describes the periodic structure as an interference pattern between the incident light and surface-scattered (e.g., plasmonic) wave \cite{Sipe}. For the excitation of surface plasmons the surface must be conductive, which is the case for rGO.
Although, the conductivity is not necessarily important for the LIPSS formation, because these structures are observed also in dielectrics like e.g., in sapphires, salts and diamonds \cite{Costache,BonseRev}.

The third scenario (simultaneous reduction and LIPSS formation) also cannot be excluded, because the hydrodynamic instabilities influence the LIPSS formation \cite{Mechanisms2016} in metals and may also be important for the GO layer. According to the literature \cite{Zhang.2010}, the reduction of graphene oxide reduces its volume. If the reduction is not homogeneous, at the sites, where the reduction is more effective, the volume (hence, the height) of the GO decreases, inducing hydrodynamic flow of the GO towards this site from the neighborhood. This will result in (1) more graphene material and (2) higher optical absorption of this area, which provides the positive feedback needed for the crests formation of the rGO LIPSS \cite{Gurevich_Levy,GOAbsorption}.

\section{Conclusion}
While the laser induced graphene oxide reduction allows to laser scribe micropatterns or circuits into the GO layer, the LIPSS can be used to generate self-organized periodic nanopatterns on the surface.
 This enables simultaneous reduction and nanopatterning of GO for rapid and flexible production of rGO structures. In particular, in this paper we demonstrated formation of LIPSS upon reduction of GO  and studied the laser parameters needed for the pattern formation and for the GO reduction. We notice that these LIPSS are different from usually observed laser-induced periodic patterns in metals, semiconductors  or dielectrics, because in our experiments the periodic pattern is not only reflected in the topography but also the chemical composition of the surface is modified. 
 
The physical mechanisms of the LIPSS formation on graphene oxide are not entirely clear. The ripples appear either {\it after} or {\it simultaneously} with the GO reduction. In the first case, the ripples are formed on a conductive rGO surface, on which surface electromagnetic waves can be excited.
In the second case
the ripples are formed on a non-conductive surface, hence the LIPPS couldn't be explained only through excitation of plasmons and their interference with the incident light. Thus, the self-organization model should contribute to their formation.

\section*{Acknowledgments}
We would like to thank the financial support of the Federal Ministry of Education and Research of Germany within the m-era.net project "CMOT-Investigation and tuning of graphene electrodes for solution-processable metal oxide thin-film transistors in the area of low-cost electronics" (03XP0014). We gratefully acknowledge the support by Vincent Layes from Experimental Physics II at the Ruhr-Universit\"at Bochum for XPS measurements, Xiao Wang affiliated with the Center for Interface-Dominated High Performance Materials (ZGH) at the Ruhr-Universit\"at Bochum for AFM measurements and Shizhou Xiao from EgdeWave GmbH for laser scanning measurements.
\section*{References}

\end{document}